\input{aipcheck}

\documentclass[
    ,final            % use final for the camera ready runs
  ]
  {aipproc}

\layoutstyle{6x9}

\def\cref#1{Chapt.\,\ref{#1}}
\def\Cref#1{Chapter~\ref{#1}}

\def\fref#1{Fig.\,\ref{#1}}

\def\lleft{\textit{left}}
\def\rright{\textit{right}}
\def\LLeft{\textit{Left}}
\def\RRight{\textit{Right}}

\def\hh{\vspace*{-0mm}}

\begin{document}

\title{Early Cosmic-Ray Work Published in German
\footnote{Invited talk, given at the Centenary Symposium 2012: Discovery of
Cosmic Rays, June 2012, Denver.}}


\author{J\"org R.\ H\"orandel}{
  address={Department of Astrophysics/IMAPP, Radboud University Nijmegen,\\
    P.O. Box 9010, 6500 GL Nijmegen, The Netherlands ~---~
    http://particle.astro.ru.nl}
}

\begin{abstract}
The article gives an overview on early cosmic-ray work, published in German
in the period from around 1910 to about 1940.
\end{abstract}

\maketitle

%%%%%%%%%%%%%%%%%%%%%%%%%%%%%%%%%%%%%%%%%%%%
%% MAINMATTER
%%%%%%%%%%%%%%%%%%%%%%%%%%%%%%%%%%%%%%%%%%%%

\section{Introduction}
The electric conductivity of air was intensely studied in the first decade of
the twentieth century \cite{elstergeitel}. This led eventually to the
discovery of cosmic rays in 1912. Traditionally, in this period many scientists
published in German language in journals like "Physikalische Zeitschrift",
"Zeitschrift f\"ur Physik", or "Naturwissenschaften".
In the following we will review early cosmic-ray work published in German from
around 1910, when the field of cosmic-ray research started, to about 1940, when
publications in German vanished due to the general political development.
The idea of this article is to make some of the early works available to
readers, who do not have access to German articles.
In the early German literature, several names are used to describe "cosmic
rays": "H\"ohenstrahlung" (high-altitude radiation), "Hesssche Strahlung" (Hess
rays), and "Ultrastrahlung" (ultra rays).
For further studies we recommend the books by V.F.\ Hess \cite{hessbook}, H.\
Geiger \cite{geigerbook}, and W.\ Heisenberg \cite{heisenbergbook}.

\section{The Beginnings}
\begin{figure}[t]
  \includegraphics[width=0.38\textwidth]{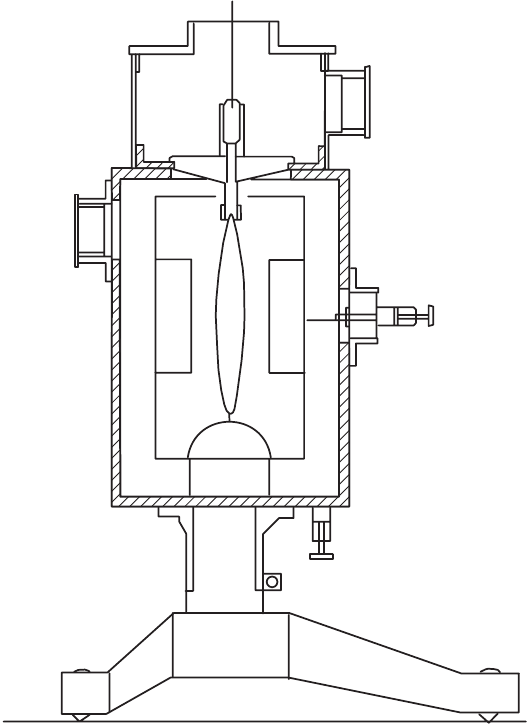}
  \includegraphics[width=0.62\textwidth]{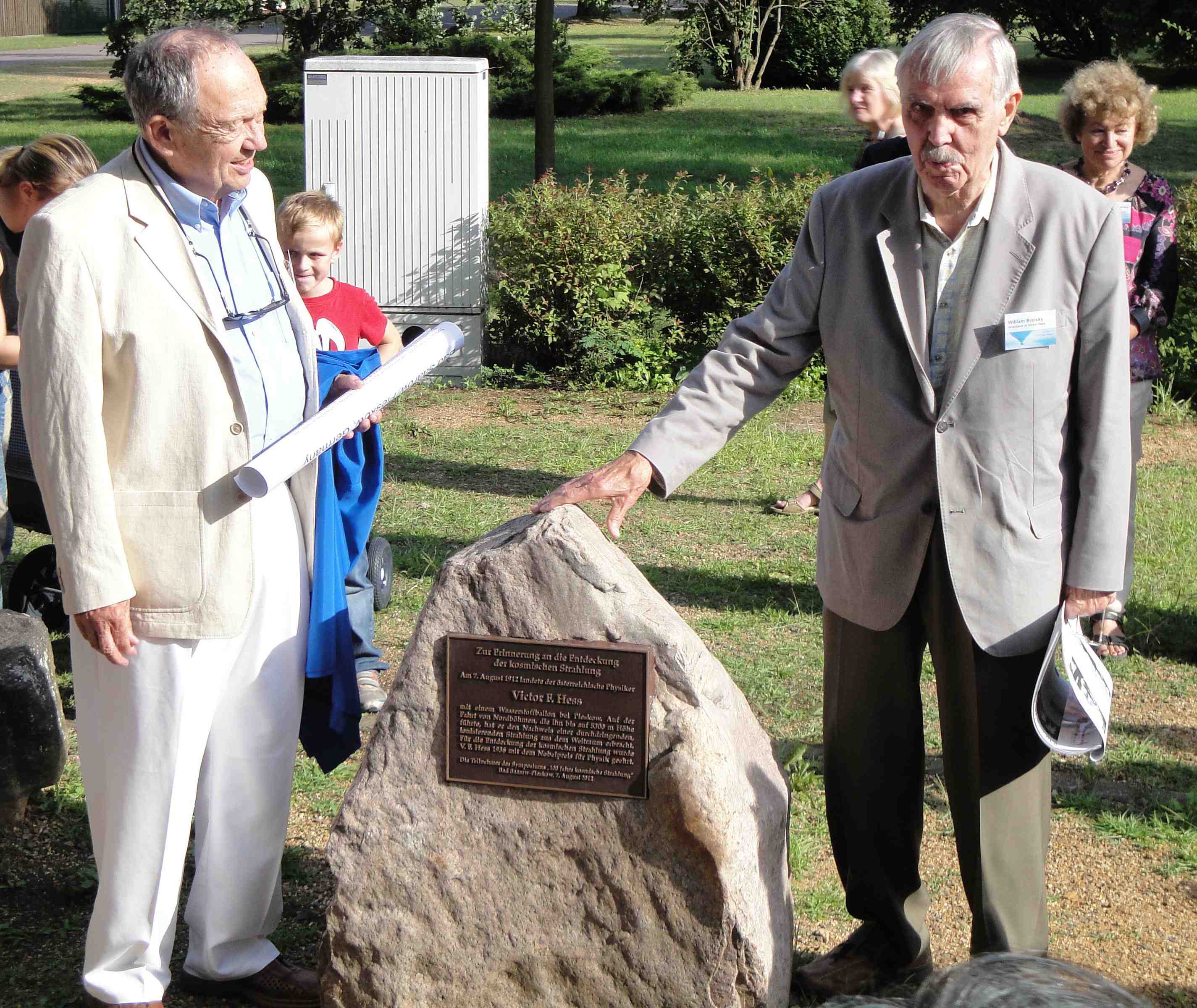}
  \caption{\LLeft: Electrometer after Th. Wulf \cite{wulfelektrometer}.
	   \RRight: Two grandsons of V.F.~Hess revealing a plaque to
  commemorate the discovery of cosmic rays on August 7th, 2012, close to the
  presumed landing site of V.F.~Hess in Pieskow close to Berlin. It reads:
 "To commemorate the discovery of cosmic rays. On 7 August 1912
 landed the Austrian physicist Victor F.\ Hess with a hydrogen balloon close to
 Pieskow. On the journey from Lower-Bohemia he reached an altitude of 5300~m and
 he proved the existence of a penetrating, ionizing radiation from outer space.
 For the discovery of cosmic rays V.F.\ Hess has been awarded the Nobel Prize in
 Physics in 1936. The participants of the symposium '100 years cosmic rays',
 Bad Saarow-Pieskow, 7 August 2012".\hh} 
  \label{fig1}
\end{figure}

In the early twentieth century the electrometer was the standard instrument to
study radioactivity and the related conductivity of air.  It was known that
radioactivity ionizes air (or gases in general) and an electrometer in the
vicinity of a radioactive source will be discharged.  One of the best
electrometer builders of this time was the Jesuit monk Theodor Wulf.  In 1909
he publishes on "A new Electrometer for static charges"
\cite{wulfelektrometer}.  A schematic view of his apparatus is given in
\fref{fig1} (\lleft).  Heart of the device is a pair of quartz fibers. They are
attached at the bottom to a further, bend quartz fiber, which acts as a spring.
By adjusting the tension on this spring, the sensitivity of the electrometer
can be adjusted.  The distance of the two fibers is measured through a
microscope, which is attached at the circumference of the device.  Series
production of the devices was provided through the company G\"unther \&
Tegetmeyer, Braunschweig.

Wulf used his apparatus e.g.\ to measure small capacities \cite{wulfcapacity}.
Main application however, was a survey to find the origin of the radioactivity
in the air. In his article "On the origin of the gamma radiation in the
atmosphere" \cite{wulf} he describes a survey, conducted in Germany, the
Netherlands, and Belgium, where he measured the intensity of the radiation in
various places. He finds an anti-correlation between the radiation intensity
and the ambient air pressure. His explanation sounds today rather exotic: one
observes less radiation at higher pressure, since the radioactive air is
pressed back into the soil/ground.\footnote{From a present point of view, in
which the pressure effect is explained due to a variation of the absorber column
density in the atmosphere, one may wonder that observing a pressure dependency
has not led to the conclusion that the radiation penetrates the atmosphere from
above.}
He summarizes his article \cite{wulf}: "The contents of this article is best
summarized as follows.  We report on experiments, which prove that the
penetrating radiation is caused by radioactive substances, which are located in
the upper layers of soil up to a depth of about 1~m. If a fraction of the
radiation originates in the atmosphere, it has to be so small, that it can not
be detected with the present apparatus."

To prove this theory, Wulf carried an electrometer to the top of the Eiffel
tower in Paris ("Observations on the radiation of high penetration power on the
Eiffel tower") \cite{wulfeiffel}.  However, his measurements were not
conclusive. At 300~m above ground he observed less radiation, but the radiation
level did not vanish completely, as expected for a purely terrestrial origin.

\section{Balloon Instruments}
The next step was to carry electrometers to higher altitudes to obtain results
beyond doubt.
Among the first scientists to conduct such measurements was A.\ Gockel,
reporting on "Measurements of the penetrating radiation during balloon
campaigns" \cite{gockel}. However, his results were not conclusive.

\begin{figure}[t]
  \includegraphics[width=0.34\textwidth]{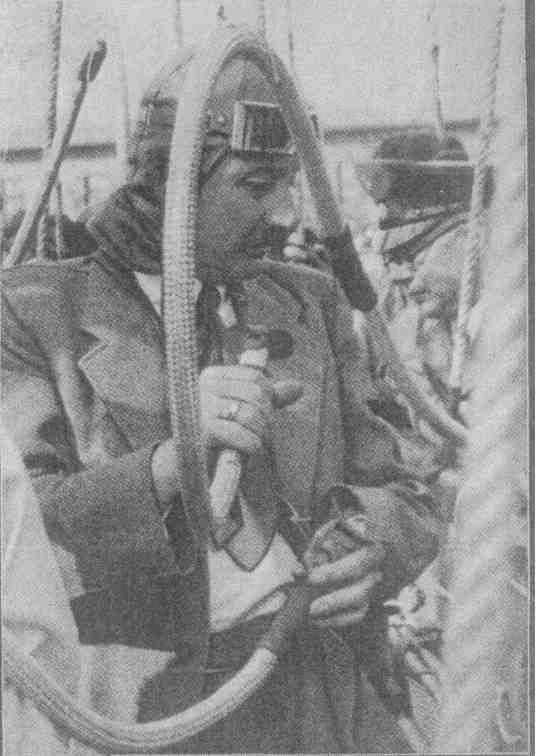}
  \includegraphics[width=0.66\textwidth]{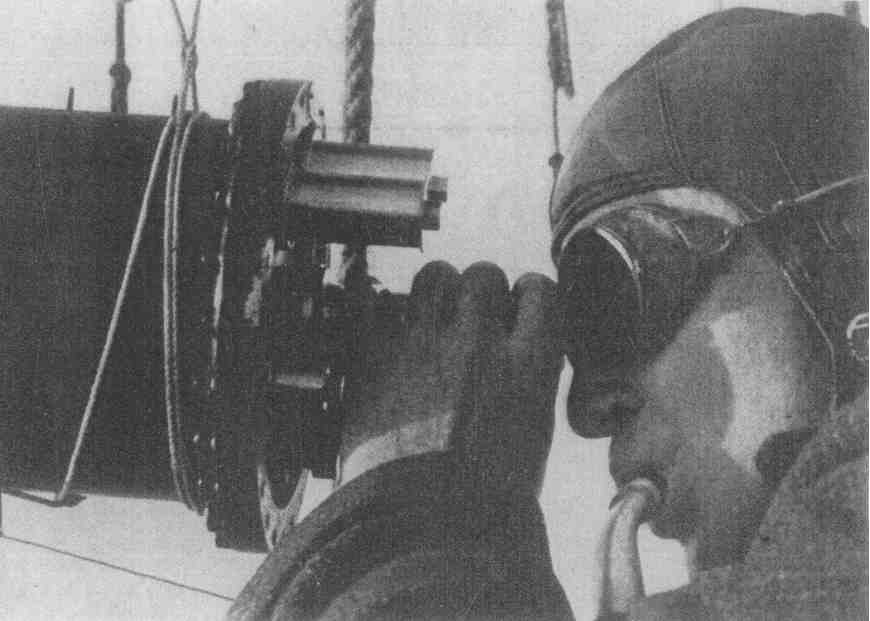}
  \caption{Dr.\ M.\ Schrenk (\lleft) with a breathing apparatus in the balloon
           gondola and V.~Masuch, reading an electrometer (\rright), preparing
           for their balloon launch on May 13th, 1934 \cite{kolhoersterunfall}.
           \hh}
  \label{fig2}
\end{figure}

The break-through has been achieved by V.F.~Hess in 1912. He used sealed,
pressure tight electrometers. Thus, the particle number density inside the
apparatus was kept constant, despite of the varying ambient temperature and air
pressure during a balloon ascent.  In his article "On the observation of the
penetrating radiation in seven free balloon campaigns" \cite{hess} Hess reports
on balloon ascends between April and July 1912.  The decisive launch was
conducted on August 7th, 1912 from Aussig an der Elbe (Austrian Empire).  At
6:12 AM the balloon "Bohemia", filled with 1680 m$^3$ hydrogen was launched,
carrying the pilot, Captain (of the K\&K Austrian army) W.~Hoffory, the
"meteorological observer" E.~Wolf, and the "air electrical observer" V.F.~Hess
to an altitude of 5300 m~a.s.l.  The balloon floated in northern direction,
towards Berlin and landed at 12:15 PM, close to the village Pieskow (Prussia),
50~km east of Berlin.  On August 7th, 2012, a plaque has been erected close to
the assumed landing site to commemorate the discovery of cosmic rays, see
\fref{fig1} (\rright).

Hess conducted measurements with three independent electrometers during the
flight.  The radiation intensity recorded by Hess as a function of altitude
exhibits first a decrease (as expected for a terrestrial origin) but then a
strong increase above 1400~m. Thus, the terrestrial origin has been disproved.
Hess summarizes his findings: "The results of the present observations can be
most likely explained through a radiation of very high penetrating power,
impinging onto the atmosphere from above, and being capable to cause the
observed ionization in closed vessels even in the lowest layers of the
atmosphere.  The intensity of the radiation exhibits timely variations on
hourly timescales.  Since I did not find a reduction of the radiation intensity
during night or during a solar eclipse, the Sun can be excluded as the origin
of this hypothetical radiation." Hess has been awarded the Nobel Prize in 1936
for his revolutionary findings.

The measurements were extended by W.~Kolh\"orster to higher altitudes.  He
conducted "Measurements of the penetrating radiation in a free balloon at high
altitudes" \cite{kolhoersterballon} to altitudes exceeding 9~km above sea
level. These observations clearly demonstrated an increase of the intensity as
a function of altitude, thus, clearly confirming an extra-terrestrial origin.
Kolh\"orster has constructed his own electrometers ("A new thread
electrometer") \cite{kolhoersterelektrometer}.

\begin{figure}[t]
  \includegraphics[width=0.6\textwidth]{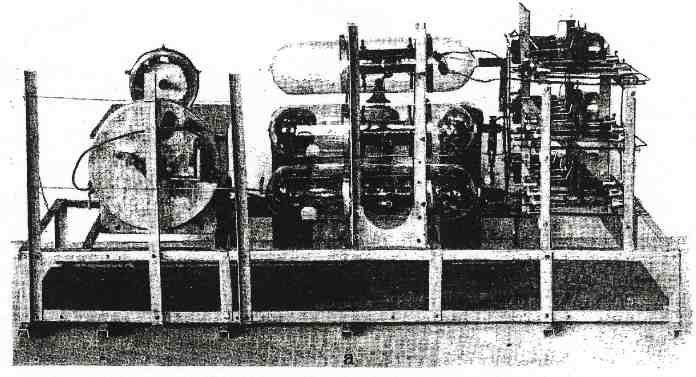}
  \includegraphics[width=0.4\textwidth]{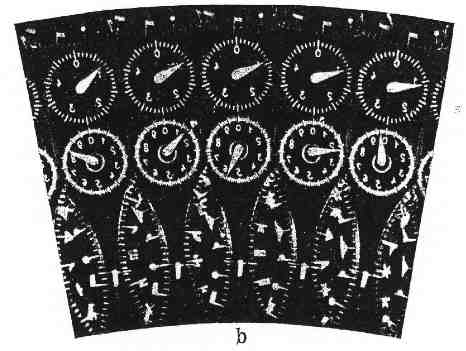}
  \caption{Recording instrument used by Pfotzer \cite{pfotzer}.
    \LLeft: the apparatus consists of (from left to right) a photographic
      recording unit, the electron vacuum tubes for the coincidence circuit,
      and a particle hodoscope, comprising of $3\times3$ Geiger-M\"uller
      tubes.
    \RRight: a photographic plate with the instrument readings, recorded
      automatically during the flight.\hh}
  \label{fig3}
\end{figure}

On May 13th, 1934 a balloon campaign was conducted under the leadership of
W.~Kolh\"orster to measure cosmic radiation up to altitudes of 12\,000~m
\cite{kolhoersterunfall}. The balloon shell was comprised of two layers of
cotton fabric, with a rubber layer in between. It had a diameter of 26.3~m and
was filled with 10\,000~m$^3$ hydrogen.  It carried a wicker basket gondola
with the dimensions 2.3~m $\times$ 1.8~m.  To be able to breath at high
altitudes, the crew used a breathing apparatus as shown in \fref{fig2}.
It is comprised of an oxygen pressure bottle, a pressure reducing
valve, and a mouth piece.  They carried an oxygen supply for four hours.  The
balloon was launched on May 13th, 1934 at 8:32 AM in Bitterfeld (Prussia).
During the balloon flight a tragic accident happened and the two collaborators
of Kolh\"orster died in the balloon gondola: Dr.~M.~Schrenk and V.~Masuch.  The
dead bodies and the balloon were found close to Sebesh (Russia) close to
midnight, about 1400~km from the launch point.

As a next step in the historical development the electroscopes were replaced by
a new type of detector: the Geiger-M\"uller tubes \cite{geigermueller}. An
essential step towards unmanned balloons with an automatic read-out of the
measurement devices was the invention of the coincidence technique, reported
by W.~Bothe and W.~Kolh\"orster in "The nature of the high-altitude radiation"
in 1929 \cite{bothekolhoerster}.  Two Geiger-M\"uller tubes have been operated
in coincidence with a metal absorber between the two tubes.  The intensity of
the penetrating radiation has been measured as function of the thickness of the
absorber material.  For the discovery of the coincidence technique, W.~Bothe
has been awarded the Nobel Prize in 1954.

\begin{figure}[t]
  \includegraphics[width=0.37\textwidth]{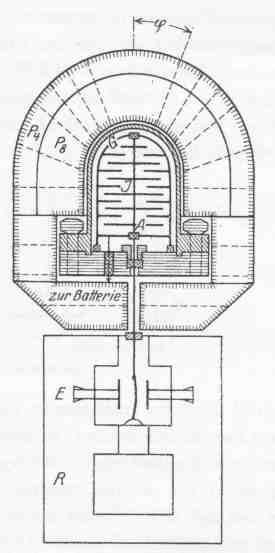}
  \includegraphics[width=0.63\textwidth]{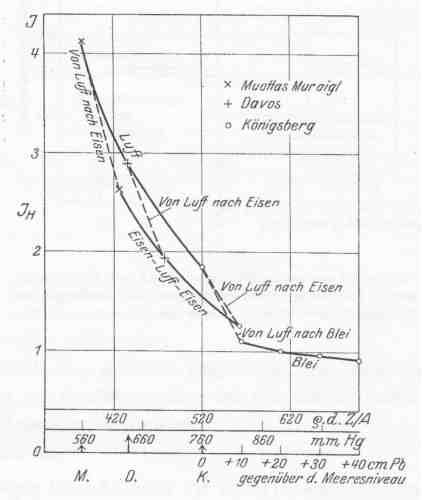}
  \caption{\LLeft: Ionization chamber with electrometer read-out, used by
      Steinke to measure the radiation intensity.
    \RRight: Measured intensity of the penetrating radiation as a
    function of the atmospheric overburden \cite{steinkeabsorption}.\hh}
  \label{fig4}
\end{figure}

G.~Pfotzer constructed a particle hodoscope, comprising of a matrix of
$3\times3$ Geiger-M\"uller tubes, operated in coincidence \cite{pfotzer}, see
\fref{fig3} (\lleft).  The coincidences were recorded with an electric circuit,
build with electron vacuum tubes.  The readings of the instruments (particle
rate, ambient air pressure and temperature) were photographed with an automatic
camera, taking pictures in a predefined time interval. After each picture, the
photographic plate was rotated. After the flight, the photographic plate had to
be recovered and it was analyzed under a microscope, see \fref{fig3} (\rright).
Pfotzer measured the particle rate as function of pressure/altitude up to a
height of 29~km.

He reports about "Three-fold coincidences of the ultra rays from vertical
direction in the stratosphere" \cite{pfotzer}. The measured intensity exhibits
a strong increase as a function of the altitude, reaching a maximum at a height
of about 15~km, where the particle rate is more than 20 times higher, as
compared to sea level.The maximum is today referred to as "Pfotzer maximum"
after its discoverer.

\begin{figure}[t]
  \includegraphics[width=0.38\textwidth]{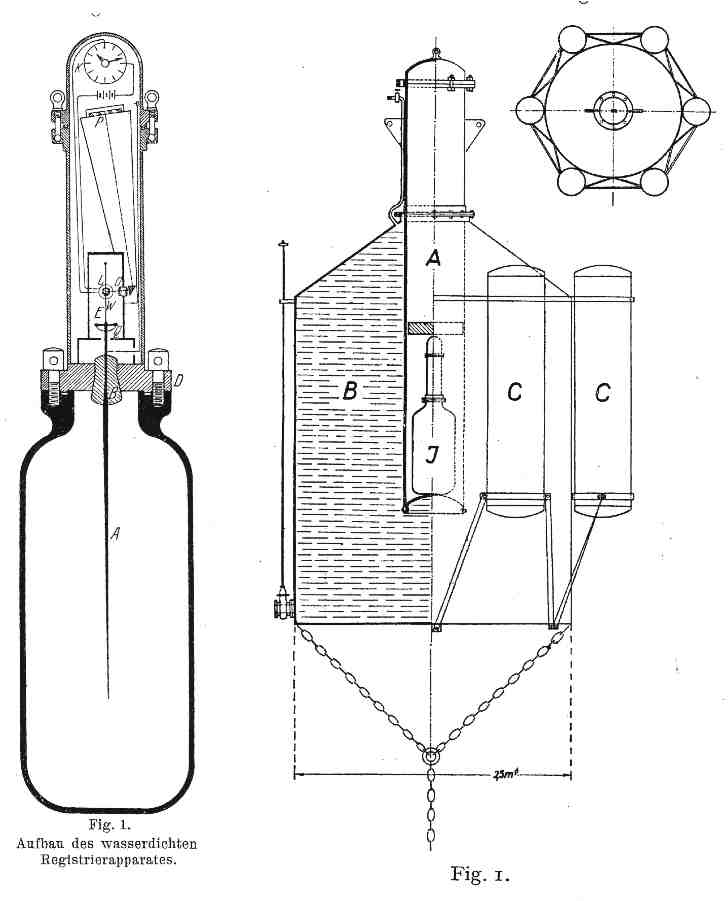}
  \includegraphics[width=0.62\textwidth]{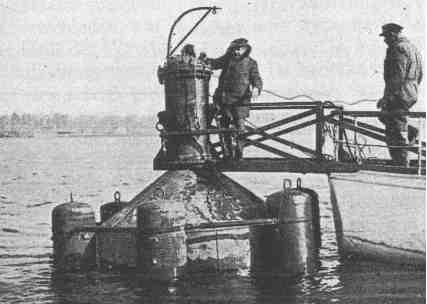}
  \caption{An ionization chamber with automatic read-out, used by Regener to
    measure the absorption of the radiation in Lake Constance
    \cite{regenerbodensee}.\hh}
  \label{fig5}
\end{figure}

\section{Absorption of Cosmic Radiation}
E.~Steinke conducted "New investigations of the penetrating Hess rays"
\cite{steinkeabsorption} and he studied the absorption of the radiation in the
atmosphere in 1928.  He used an ionization chamber, read out by a Wulf
one-string electrometer.  The ionization chamber was shielded by a segmented
12~cm iron absorber. The set-up is shown in \fref{fig4} (\lleft).  Individual
segments of the shield could be removed, thus, the apparatus was sensitive to
radiation from a certain zenith angle interval.  He measured the radiation
intensity as a function of the zenith angle at different altitudes above sea
level, starting from K\"onigsberg (Prussia) 0~m a.s.l., via Davos (Switzerland)
1600~m a.s.l., to Muottas Muraigl 2500~m a.s.l.  The results are depicted in
\fref{fig4} (\rright).  It should be noted that he already put on the abscissa
$d\cdot \rho Z/A$, i.e.\ $\Delta E=d\cdot dE/dx$, the energy loss in the
atmosphere, measured from the top.\footnote{H.~Bethe published his "Theory of
the passage of fast corpuscular radiation through matter" in 1930
\cite{bethededx}.  The Bethe-Bloch equation to describe the energy loss of a
particle traversing matter gives the proportionality $dE/dx\propto\rho Z/A$.}  

Steinke points out that "the direction and absorption measurements allow a
flawless separation of the Hess rays from the ambient radiation. The angular
distribution of the Hess rays, corresponds to values, which are compatible with
the assumption of an isotropic radiation from outer space, taking into account
the absorption along the different pathlengths through the atmosphere."
He points out that the absorption coefficient depends on the absorber material,
and not only on the column density, see \fref{fig4} (\rright) and he suggests to
describe the radiation with two components: a "hard" and a "soft"
component.\footnote{Similar measurements were conducted later by B.\ Rossi and
colleagues at Mt.\ Evans in Colorado and Rossi realized that the differences in
the absorption curves are caused by the decay of muons \cite{rossi}, p.\ 118.}

Steinke conducted systematic studies of the intensity variations. He reports
"On variations and the barometric effect of cosmic ultra rays at sea level"
\cite{steinkebarometric} and describes periodic and non-periodic variations,
such as an anti-correlation between the ambient pressure and the radiation
intensity (barometric effect), as well as an annual modulation and a sidereal
modulation of the cosmic-ray intensity.

\begin{figure}[t]
  \includegraphics[width=0.7\textwidth]{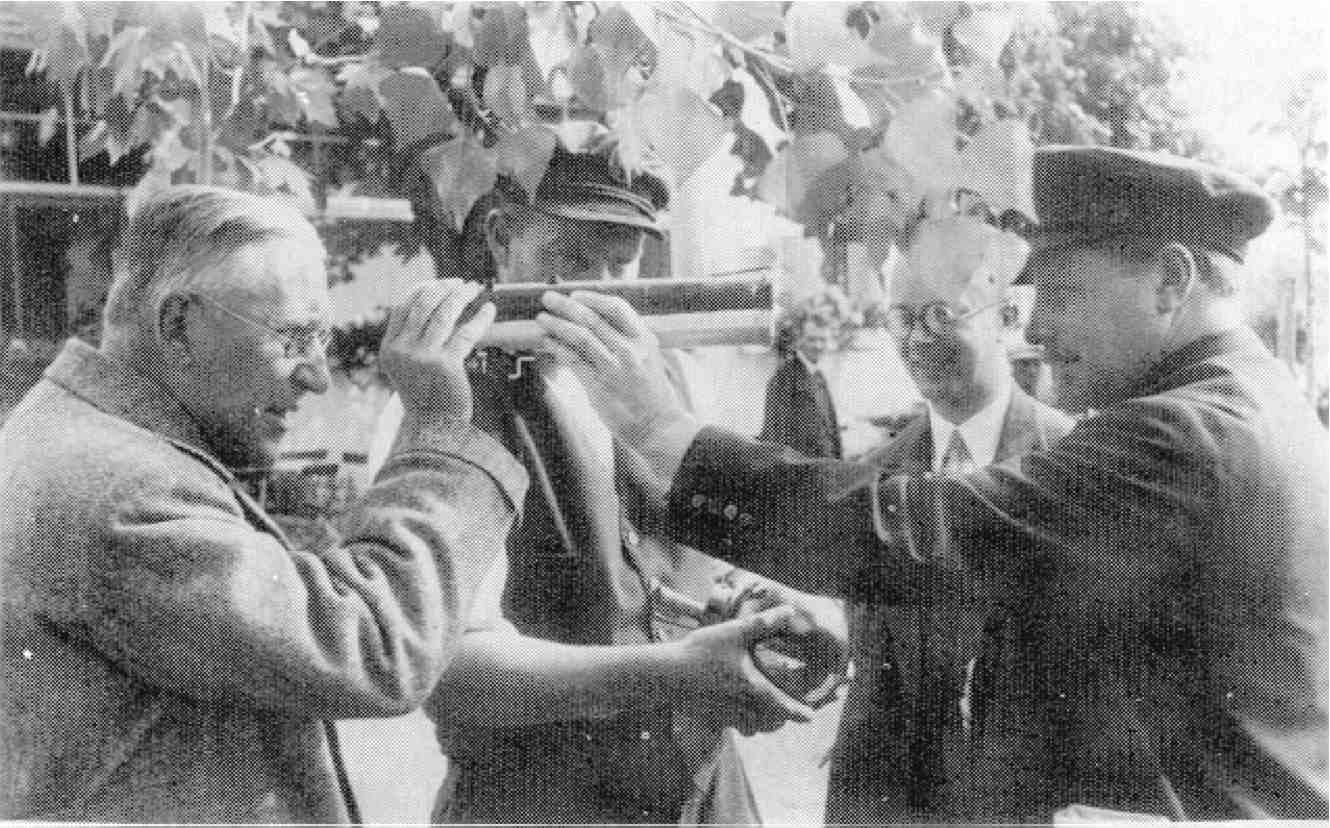}
  \caption{Three pioneers of cosmic-ray research: Regener (right) demonstrates
	   his balloon electrometer to Hess (left) and Steinke (center).
           Immenstaad/Lake Constance, August 1932.\hh}
  \label{fig7}
\end{figure}

To study the absorption of the ultra rays in water E.~Regener constructed an
ionization chamber with electrometer read-out. The apparatus was attached to a
buoy, as sketched in \fref{fig5}, and could be lowered into the water
to record the radiation intensity.  The apparatus recorded automatically the
intensity every hour for up to eight days. Regener conducted measurements in
Lake Constance up to a depth of 250~m.  He reports on "The absorption curve of
the ultra radiation and its interpretation" \cite{regenerbodensee} and
discusses the attenuation of the ultra rays, measured in meter water
equivalent, counted from the top of the atmosphere.
He states that cosmic rays are with high probability undulatory radiation and
he strictly denies a corpuscular nature of cosmic rays.
As a nice anecdote it may be remarked that Regener named the boat, which was
used to conduct the measurements, "Undula", indicating his believe 
about the nature of the radiation.
During the measurements at Lake Constance the photograph shown in \fref{fig7}
has been taken.

It was also Regener, who pointed out in his article on "The energy flux of the
ultra rays" \cite{regeneredensity} in 1933 that the energy flux of cosmic rays
corresponds roughly to the energy flux of starlight.  For the cosmic-ray
intensity at the top of the atmosphere he gives a value of
$3.53\cdot10^{-3}$~erg\,cm$^{-2}$\,s$^{-1}$ and he notices that this value
corresponds to a flux of a couple of hundred $\alpha$-particles per cm$^2$ and
s, impinging onto the atmosphere of the Earth.  He also conducts an interesting
calculation: a celestial body, which is exposed to cosmic rays will be heated
through the absorption of cosmic rays. The body reaches a temperature of about
2.8~K.

\section{Extensive Air Showers and Nuclear Interactions}
\begin{figure}[t]
  \includegraphics[width=0.53\textwidth]{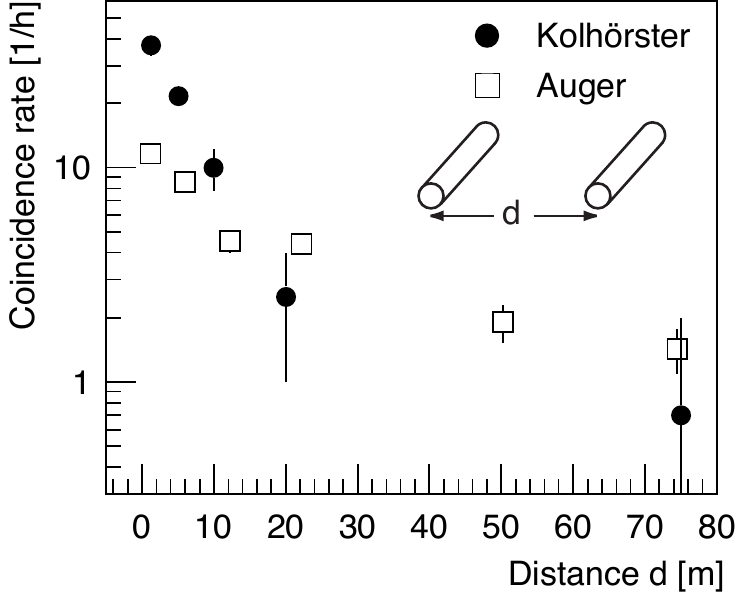}
  \includegraphics[width=0.47\textwidth]{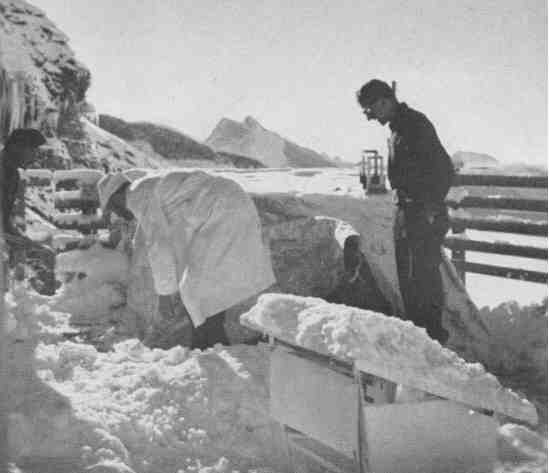}
  \caption{\LLeft: Coincidence rate as a function of the distance between two
	   Geiger-M\"uller counters as obtained by W.~Kolh\"orster
           \cite{kolhoersterschauer} and P.~Auger \cite{augerschauer}.
	  \RRight: P.~Auger measuring air showers at the Jungfraujoch in
          Switzerland \cite{augerbuch}.\hh}
  \label{fig6}
\end{figure}

The coincidence technique also brought the next step forward in the observation
of cosmic particles at ground level.  W.~Kolh\"orster placed two
Geiger-M\"uller counters next to each other and operated them in coincidence.
He recorded the number of coincidences as a function of the distance between
the two counters, as illustrated in \fref{fig6} (\lleft).  Objective of these
investigations was to determine the random coincidence rate between the two
counters. However, the measurements clearly indicated an excess of
coincidences, which led to the discovery of extensive air showers.  In 1938 he
reports about "Coupled high-altitude rays" \cite{kolhoersterschauer}.  He
explains that the observed particles are secondary particles from cosmic rays,
i.e.\ air showers. The secondary particles in the showers are produced at high
altitudes above the ground and they are distributed over a large area on the
ground. Coincidences have been registered even up to a distance of 75~m.  To
prove his findings he also conducted an additional experiment, using three
counters, operated in a three-fold coincidence with a good time resolution of
$5~\mu$s.  Also with this set-up he found an excess of coincidences, clearly
confirming his discovery of air showers.

Similar investigations were conducted by P.~Auger on the Jungfraujoch in the
Swiss Alps, see \fref{fig6} (\rright), and at other places \cite{augerschauer}.
The results obtained by Auger are in good agreement with the measurements by
Kolh\"orster and colleagues as can be inferred from \fref{fig6} (\lleft).

However, the physical interpretation of the measured showers was not easy.  The
development of electromagnetic cascades has been known (e.g.\
\cite{Bethe:1934za} and later written up in \cite{heitlerbook}). But without
knowing the pion (discovered in 1947 \cite{dispion}) and the development of
hadronic showers, it was hard to fully understand the measured attenuation
coefficients.  A first step towards this was the discovery of hadronic
interactions.

\begin{figure}[t]
  \includegraphics[width=0.54\textwidth]{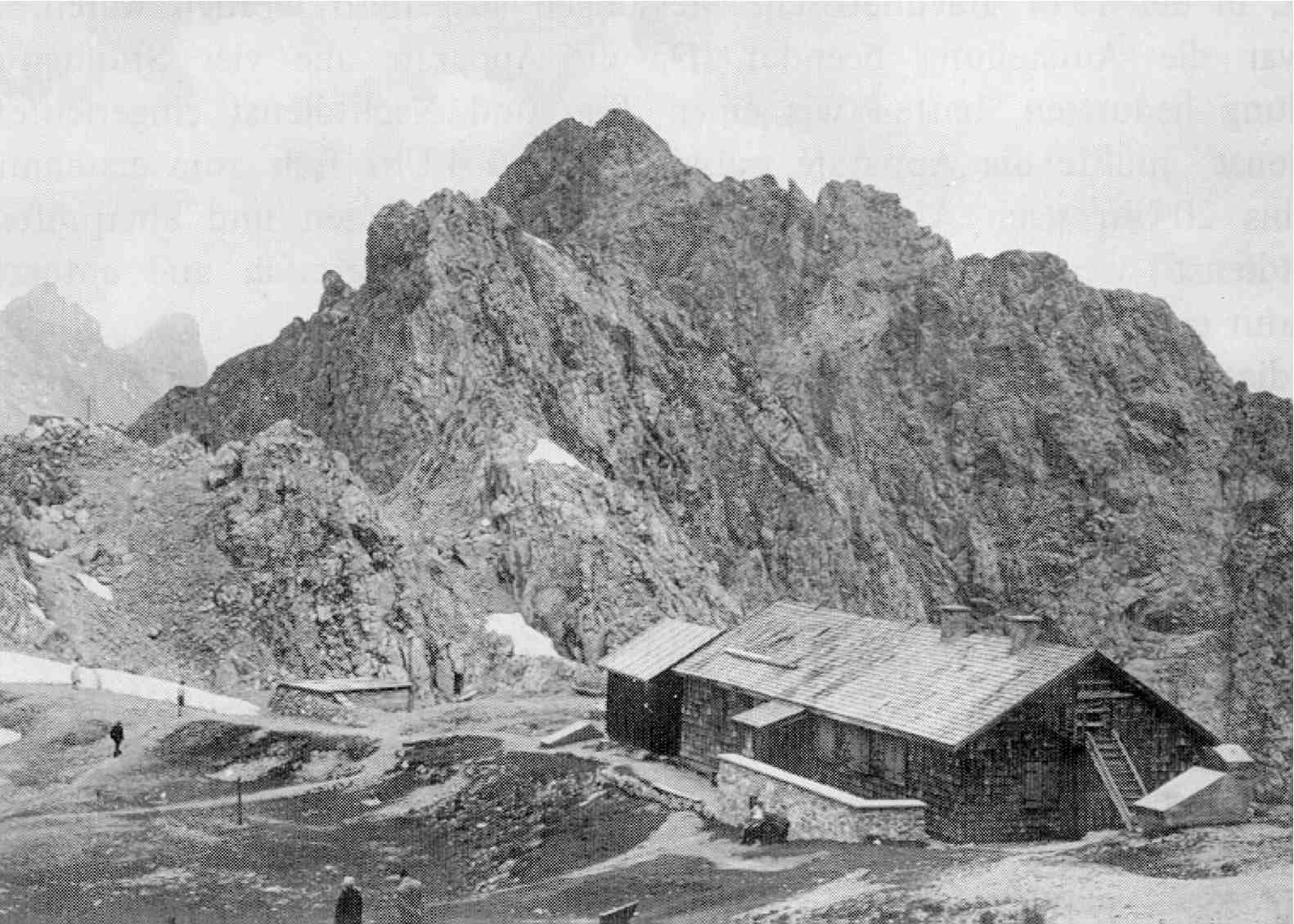}
  \includegraphics[width=0.46\textwidth]{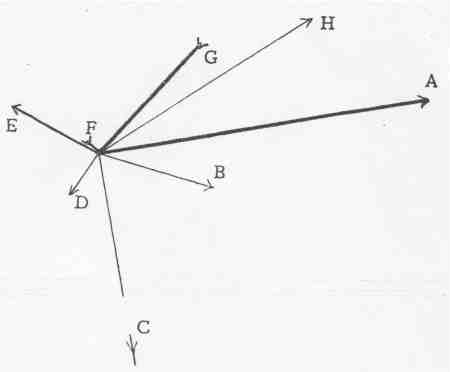}
  \caption{\LLeft: High-altitude laboratory to study cosmic rays and their
    interactions, established by V.~Hess on the Hafelekar mountain
    \cite{hessbook}.
    \RRight: A hadronic interaction process (a "star") recorded in a
     photographic emulsion \cite{blauwambacher}.\hh}
  \label{fig8}
\end{figure}

V.~Hess has established a high-altitude laboratory to study cosmic rays at the
Hafelekar mountain at an altitude of 2300~m a.s.l.\ above Innsbruck in Austria.
The laboratory is shown in \fref{fig8} (\lleft).  M.~Blau and H.~Wambacher
worked in this laboratory and they used photographic plates to investigate
cosmic rays and their interactions.  They studied "Disintegration processes by
cosmic rays with the simultaneous emission of several heavy particles"
\cite{blauwambacher}.  They found hadronic interactions of cosmic particles
with the nuclei inside the photographic emulsion. Such a "star" is depicted in
\fref{fig8} (\rright). Such processes were interpreted as the disintegration of
an atom(ic nucleus) in the emulsion.  This was the birth of the emulsion
technique to study the interactions of particles.  

%%%%%%%%%%%%%%%%%%%%%%%%%%%%%%%%%%%%%%%%%%%%%%%%
%% BACKMATTER
%%%%%%%%%%%%%%%%%%%%%%%%%%%%%%%%%%%%%%%%%%%%%%%%

\begin{theacknowledgments}
 The author thanks Jonathan Ormes for organizing the wonderful
 symposium to celebrate the 100th anniversary of the discovery of cosmic rays.
\end{theacknowledgments}

\bibliographystyle{aipproc}   % if natbib is available

%\bibliography{cr}

\end{document}